\documentstyle[aps,floats,psfig]{revtex}

\begin{document}
\draft 

\wideabs{

\title{Edge and bulk electron states in a quasi-one-dimensional metal
in a magnetic field: The semi-infinite Wannier-Stark ladder}

\author{Victor M.~Yakovenko\cite{Yakovenko} and Hsi-Sheng Goan}

\address{Department of Physics and Center for Superconductivity,
University of Maryland, College Park, MD 20742-4111}

\date{{\bf cond-mat/9706274}; v.1 June 26, 1997; v.2 March 9, 1998;
     v.3 June 12, 1998}

\maketitle

\begin{abstract}
  
  We study edge and bulk open-orbit electron states in a
  quasi-one-dimensional (Q1D) metal subject to a magnetic field. For
  both types of the states, the energy spectrum near the Fermi energy
  consists of two terms. One term has a continuous dependence on the
  momentum along the chains, whereas the other term is quantized
  discretely. The discrete energy spectrum is mathematically
  equivalent to the Wannier-Stark energy ladder of a semi-infinite 1D
  lattice in an effective electric field. We solve the latter problem
  analytically in the semiclassical approximation and by numerical
  diagonalization.  We show explicitly that equilibrium electric
  currents vanish both at the edges and in the bulk, so no orbital
  magnetization is expected in a Q1D metal in a magnetic field.

\end{abstract}

}

\section{Introduction}
\label{sec:intro}

In a strong magnetic field, the quasi-one-dimensional (Q1D) organic
conductors of the (TMTSF)$_2$X family \cite{TMTSF} (also known as the
Bechgaard salts) exhibit very interesting phenomena, such as magnetic
oscillations, magnetic-field-induced spin-density wave (FISDW), and
the quantum Hall effect (QHE) (see, for example, review
\cite{Chaikin96}).  Because the Fermi surfaces of Q1D metals are open,
these phenomena have different mechanisms in Q1D conductors compared
to more conventional materials with closed Fermi surfaces.  For
example, the QHE exists only in the FISDW state, but not in the
metallic state of Q1D conductors \cite{Lederer96,Yakovenko96}.  Thus
far, the theory of the Bechgaard salts focused mostly on the bulk
electron properties (see, for example, review \cite{Lederer96}). Only
recently the edge aspects of the QHE in the Bechgaard salts attracted
attention \cite{Chalker95,MPAFisher96}. An explicit picture of the QHE
in the FISDW state in terms of the edge states was developed in Ref.\ 
\cite{Yakovenko96}. However, that work did not take into account
possible deformations of the electron wave functions near the edges.
In the current paper, we present a detailed study of the electron wave
functions and energies near the edge of a Q1D conductor in the
metallic (not FISDW) state. This work may serve as a starting point
for a more accurate theory of the edge states in the FISDW state and
their role in the QHE. Proper description of the edge states is also
important for the theory of the cyclotron resonance in Q1D metals
\cite{Lebed}.

The edge states of electrons in a Q1D metal in a magnetic field were
studied semiclassically by Azbel and Chaikin \cite{Azbel87,Azbel-1990}
and numerically by Osada and Miura \cite{Osada89}. In Ref.\ 
\cite{Azbel87} the WKB quantization condition was applied to the
problem inconsistently, which resulted in a wrong conclusion that the
edge states have a discrete energy spectrum, whereas the bulk states
have a continuous one. This statement was also repeated in Ref.\ 
\cite{Azbel-1989}. It was claimed in Refs.\ \cite{Azbel87,Azbel-1990}
that the electron edge states produce thermodynamic oscillations of
magnetization in a Q1D metal with an open Fermi surface.  In the
present paper, we clear up the confusion and show that the energy of
either a bulk or an edge state is a sum of two terms, one of which has
a continuous spectrum and the other discrete (in the approximation
where the longitudinal electron dispersion law is linearized near the
Fermi energy). The WKB quantization condition determines the discrete
energy terms of both the edge and the bulk states. In Appendix, we
explicitly point out a mathematical error in Ref.\ \cite{Azbel-1990}
that led to the wrong conclusions. In Sec.\ \ref{sec:currents}, we
show explicitly that the equilibrium electric currents vanish both at
the edges and in the bulk, so no orbital magnetization is expected in
a magnetic field. This result is in agreement with independence of the
bulk internal energy of a Q1D metal with an open Fermi surface on a
magnetic field \cite{LAK}.

The Schr\"odinger equation that we solve analytically (Sec.\ 
\ref{sec:analytic}) and numerically (Sec.\ \ref{sec:numerical}) in
order to find the discrete part of the electron energy is
mathematically equivalent to the equations that describe the
Wannier-Stark ladder \cite{Wannier} of a semi-infinite 1D lattice in a
uniform electric field.  An analytical solution of this problem in
terms of special functions was obtained in Refs.\ 
\cite{Fukuyama73,Stey-1973}, but our WKB solution is more general. Our
results might be useful for interpreting experiments on finite-size
GaAs-Ga$_{1-x}$Al$_x$As superlattices in an electric field
\cite{Mendez-1993}.

\section{Analytical solution}
\label{sec:analytic}

We model the Bechgaard salts by a 2D system that consists of 1D chains
parallel to the $x$ axis and spaced at a distance $b$, their coordinates
being $y=nb$, where $n$ is an integer number. The Fermi surface of 1D
electron motion along the chains consists of the two Fermi points
characterized by the Fermi momenta $\pm P_F$. The energy dispersion law
of the longitudinal electron motion can be linearized in the vicinity of
the Fermi energy: $\varepsilon_\|=\pm v_Fp_x$, where $v_F$ is the Fermi
velocity, the energy $\varepsilon_\|$ is counted from the Fermi energy,
and the longitudinal momenta $p_x$ are counted from $\pm P_F$ for the
two Fermi points: $p_x=P_x\mp P_F$. In this paper, we consider only the
electron states in the vicinity of the $+P_F$ Fermi point. The formulas
for the $-P_F$ electrons can be obtained by changing the sign of $v_F$.
The chains are coupled in the $y$ direction by the electron tunneling
amplitude $t$. The magnetic field $H$ is applied in the $z$ direction.
Choosing the Landau gauge, $A_x=-Hy$ and $A_y= A_z=0$, we introduce the
magnetic field into the Hamiltonian via the substitution $p_x\rightarrow
p_x-eA_x/c$, where e is the electron charge and $c$ is the speed of
light. An energy eigenfunction of electron has the factorized form:
\begin{equation}
\psi_{p_x,M}(x,n)=e^{ip_x x/\hbar}\phi_M(n).
\label{psi}
\end{equation}
The eigenfunctions of transverse motion, $\phi_M(n)$, are labeled by the
discrete quantum number $M$ and obey the following 1D discrete
Schr\"odinger equation:
\begin{equation}
n\Omega\phi_M(n)-t[\phi_M(n+1)+\phi_M(n-1)]=E_M\phi_M(n),
\label{discrete}
\end{equation}
where $\Omega$ is the characteristic energy of the magnetic field:
\begin{equation}
\Omega=ebHv_F/c.
\label{E_H}
\end{equation}
Eq.\ (\ref{discrete}) also describes a 1D lattice in the uniform
electric field $-Hv_F/c$ in the $y$ direction. This electric field
would appear in the reference frame moving with the Fermi velocity
$v_F$ due to the Lorentz transformation of the magnetic field $H$. The
energy $\varepsilon(p_x,M)$ of eigenfunction (\ref{psi}) is the sum of
the longitudinal and transverse terms:
\begin{equation}
\varepsilon(p_x,M)=v_Fp_x + E_M.
\label{E}
\end{equation}
We assume that $H$ is not too strong: $\Omega\leq2t$. The opposite
case $\Omega\geq2t$, easily treated by perturbation theory in the small
parameter $2t/\Omega$, requires unrealistically high magnetic fields
in the Bechgaard salts.

We consider a crystal that is infinite in the $x$ direction and
semi-infinite in the positive $y$ direction. The wave functions
$\phi_M(n)$ are defined at $n\geq1$ with the free boundary condition at
$n=1$. As one can see from Eq.\ (\ref{discrete}), this formulation is
equivalent to considering $\phi_M(n)$ at both positive and negative $n$
with the zero boundary condition at $n=0$:
\begin{equation}
\phi_M(0)=0.
\label{n=0}
\end{equation}

We closely follow Ref.\ \cite{Landau} in our treatment of the problem.
To solve Eq.\ (\ref{discrete}), we express $\phi_M(n)$ in terms of its
Fourier transform $\varphi_M(k)$:
\begin{equation}
\phi_M(n)=\int e^{ink}\varphi_M(k)\,\frac{dk}{2\pi}.
\label{phiMy}
\end{equation}
Eq.\ (\ref{phiMy}) defines the function $\phi_M(n)$ of the continuous
variable $n$, which has physical meaning only at the integer positive
points. The integration in Eq.\ (\ref{phiMy}) proceeds along a certain
contour in the complex plane of $k$. Eq.\ (\ref{phiMy}) satisfies Eq.\
(\ref{discrete}) provided $\varphi_M(k)$ vanishes at the ends of the
contour and obeys the following equation:
\begin{equation}
i\Omega\,\partial\varphi_M(k)/{\partial k}=
(E_M+2t\cos k)\,\varphi_M(k).
\label{differential}
\end{equation}
Solution of Eq.\ (\ref{differential}) is
\begin{eqnarray}
&&\varphi_M(k)=\exp\left[-i\int_0^{k}\xi(k',E_M)\,dk'\right]
\label{varphiMy} \\
&&=\exp[-i(E_Mk+2t\sin k)/\Omega],
\label{phase}
\end{eqnarray}
where the function
\begin{equation}
\xi(k,E_M)=[E_M-\varepsilon_\perp(k)]/\Omega
\label{Y}
\end{equation}
is defined for a general transverse dispersion law
$\varepsilon_\perp(k)$, whereas Eq.\ (\ref{phase}) is specific to
$\varepsilon_\perp(k)=-2t\cos k$. 

\begin{figure}
\centerline{\psfig{file=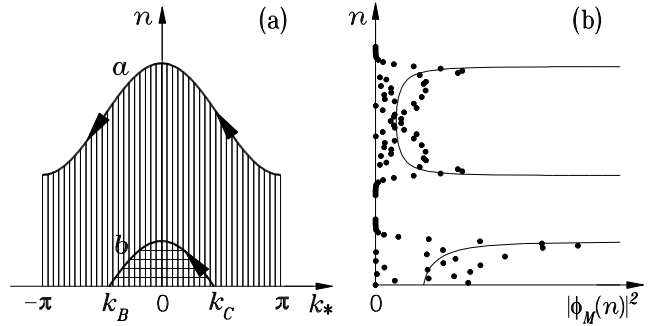,width=\linewidth,angle=-90,clip=}}
\caption{ (a) The bulk (curve $a$) and the edge (curve $b$) classical
trajectories of electrons in the phase space $(n,k_*)$. The coordinate
$n$ is confined between $(E_M\pm t)/\Omega$ for the bulk state and
between $(E_M+t)/\Omega$ and 0 for the edge state. (b) Solid lines:
Classical probability distributions for the two trajectories shown in
panel (a).  Dots: Quantum probability distributions $|\phi_M(n)|^2$ of
the two wave functions corresponding to the two trajectories. }
\label{classical}
\end{figure}

When $t\gg\Omega$, integral (\ref{phiMy}) with $\varphi_M(k)$ from Eq.\
(\ref{varphiMy}) can be taken by the method of steepest descent in the
vicinity of the points $k_*$ where the derivative in $k$ of the phase
of the integrand vanishes:
\begin{equation}
n=\xi(k_*,E_M)=(E_M+2t\cos k_*)/\Omega.
\label{star}
\end{equation}
Eq.\ (\ref{star}) can be interpreted as the classical conservation law
of the kinetic, $-2t\cos k_*$, and potential, $n\Omega$, energies of
electron. If the coordinate $n$ belongs to the classically allowed
region $[(E_M-2t)/\Omega,(E_M+2t)/\Omega]$, then $k_*$ is real; otherwise,
$k_*$ is complex. Real solutions of Eq.\ (\ref{star}) describe classical
electron trajectories in the phase space $(n,k_*)$. When $E_M>2t$, the
trajectory lies entirely within the region $n>0$ and does not cross the
boundary of the crystal at $n=0$ (curve $a$ in Fig.\
\ref{classical}(a)). When $-2t<E_M<2t$, the trajectory reaches the edge
(curve $b$ in Fig.\ \ref{classical}(a)). These two types of classical
trajectories correspond to the bulk and the edge quantum states of
electrons. The classical motion is periodic both for the bulk
trajectory, because the end points $k_*=\pm\pi$ correspond to the same
state, and for the edge trajectory, because elastic reflection at point
$k_B$ reverses the sign of $k_*$ and transfers electron back to point
$k_C$. Thus, we expect the WKB quantization condition to apply in both
cases:
\begin{equation}
\int \xi(k,E_M)\,dk=2\pi(M+\gamma),
\label{WKB}
\end{equation}
where $-1<\gamma\leq0$ is a constant, and the integral represents the
phase space area enclosed by the classical trajectory. For the bulk
and the edge trajectories $a$ and $b$ in Fig.\ \ref{classical}(a),
these areas are shaded vertically and horizontally. Contrary to Ref.\ 
\cite{Azbel87}, we find well-defined WKB quantization areas for both
the bulk and the edge trajectories.

\begin{figure}
\centerline{\psfig{file=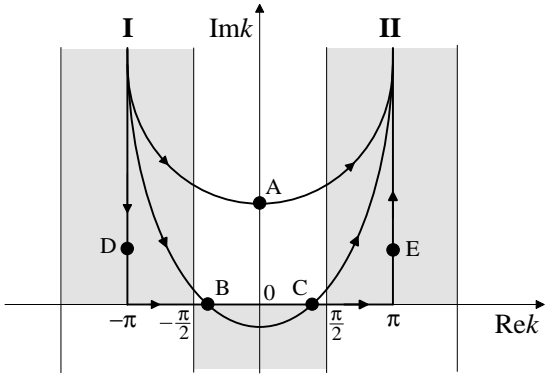,width=\linewidth,angle=90,clip=}}
\caption{ Complex plane of $k$. Thick lines show the contours of
integration in Eq.\ (\ref{phiMy}) for three different positions of the
coordinate $n$.}
\label{complex}
\end{figure}

To derive quantization condition (\ref{WKB}) for our model formally and
to find the constant $\gamma$, we need to apply the boundary conditions
properly. Integral (\ref{phiMy}) with $\varphi_M(k)$ given by Eq.\
(\ref{phase}) converges only if the ends of the integration contour
extend to infinity within the shaded areas in Fig.\ \ref{complex}, where
${\rm Im}\sin k<0$, and $\varphi_M(k)$ tends to zero at infinity. The
right boundary condition in real space, $\phi_M(n)\rightarrow0\;\;{\rm
at}\;\;n\rightarrow+\infty,$ is satisfied provided the contour of
integration starts in area $I$ and ends in area $II$ in Fig.\
\ref{complex}. Indeed, in the classically inaccessible region
$n\rightarrow+\infty$, solutions of Eq.\ (\ref{star}) are imaginary. One
of them, $k_A=i\,{\rm arccosh}(n'\Omega/2t)$, where $n'=n-E_M/\Omega$, is
represented by point $A$ in Fig.\ \ref{complex}. The contour of
integration connects regions $I$ and $II$ by passing through point $A$.
Taking integral (\ref{phiMy}) in the vicinity of point $A$ along the
direction of steepest descent, which is parallel to the real axis of $k$
in this case, we find:
\begin{equation}
\phi_M(n)\approx\exp\{-n'[\ln(n'\Omega/t)-1]\}/\sqrt{2\pi n'},
\label{phi+infty}
\end{equation}
which does satisfy the right boundary condition.

Now let us calculate $\phi_M(n)$ in the classically accessible region.
In this case, solutions of Eq.\ (\ref{star}), represented by points
$B$ and $C$ in Fig.\ \ref{complex}, are real:
$k_C=-k_B=\arccos(n'\Omega/2t)$. The contour of integration connects
regions $I$ and $II$ by passing through points $B$ and $C$:
\begin{eqnarray}
\phi_M(n)&=&\left(e^{ink_B-i\pi/4} + e^{ink_C+i\pi/4
-i\int_{k_B}^{k_C} \xi(k',E_M)\,dk'}\right) \nonumber \\
&& \times \varphi_M(k_B)/\sqrt{2\pi}\sqrt[4]{(2t/\Omega)^2-n'^2}.
\label{edge0}
\end{eqnarray}
The factors $\exp(\mp i\pi/4)$ appear in Eq.\ (\ref{edge0}), because
the directions of steepest descent for points $B$ and $C$ are at the
angles $\mp\pi/4$ to the real axis of $k$. The integral from $k_B$ to
$k_C$ in Eq.\ (\ref{edge0}) reflects the change of function
(\ref{varphiMy}) between points $B$ and $C$. For the edge states, the
point $n=0$ is classically accessible. To satisfy the left boundary
condition (\ref{n=0}), the first line in Eq.\ (\ref{edge0}) must
vanish at $n=0$.  This generates quantization condition (\ref{WKB})
with $\gamma=-1/4$ for the edge states:
\begin{equation}
\int_{-\arccos(-E_M/2t)}^{\arccos(-E_M/2t)}
\xi(k,E_M)\,dk=2\pi\left(M-\frac14\right).
\label{WKBedge}
\end{equation}
Substituting Eq.\ (\ref{Y}) into Eq.\ (\ref{WKBedge}) gives a
transcendental equation on $E_M$, which has the following explicit
solution for the states on the very edge with $M\ll t/\Omega$:
\begin{equation}
E_M=t\{-2+[(3\pi\Omega/2t)(M-1/4)]^{2/3}\}.
\label{EMedge}
\end{equation}
The total number of the edge states is $n_{\rm edge}=2t/\Omega$. Eqs.\
(\ref{WKBedge}) and (\ref{EMedge}) are similar to the edge states
quantization equations for a closed Fermi surface \cite{Prange68}.

In the classically inaccessible region $n<(E_M-2t)/\Omega$, solutions of
Eq.\ (\ref{star}), represented by points $D$ and $E$ in Fig.\
\ref{complex}, are complex: $k_{D,E}=\mp\pi+i\,{\rm
arccosh}(-n'\Omega/2t)$. The contour of integration connects regions $I$
and $II$ by passing through points $D$ and $E$:
\begin{eqnarray}
&&\phi_M(n)=(-i)\left[1-\exp\left(-i\int_{k_D}^{k_E} \xi(k',E_M)\,dk'
  \right)\right] \nonumber \\
&&\times\exp\{-n'[{\rm arccosh}(-n'\Omega/2t)+i\pi]\} \nonumber \\
&&\times \exp[-\sqrt{n'^2-(2t/\Omega)^2}]/
  \sqrt{2\pi}\,\sqrt[4]{n'^2-(2t/\Omega)^2}.
\label{bulk0}
\end{eqnarray}
The integral between $k_D$ and $k_E$ in Eq.\ (\ref{bulk0}) proceeds
along the horizontal line $[-\pi,\pi]$ and the vertical lines
$[k_D,-\pi]$ and $[\pi,k_E]$ (see Fig.\ \ref{complex}); however, the
integrals along the vertical lines cancel. To satisfy the left
boundary condition, Eq.\ (\ref{n=0}) for a semi-infinite crystal or
$\phi_M(n)\rightarrow0$ at $n\rightarrow-\infty$ for an infinite one,
the first line in Eq.\ (\ref{bulk0}) must vanish. This generates
quantization condition (\ref{WKB}) with $\gamma=0$ for the bulk
states:
\begin{equation}
\int_{-\pi}^{\pi} \xi(k,E_M)\,dk=2\pi M.
\label{WKBbulk}
\end{equation}
Substituting Eq.\ (\ref{Y}) into Eq.\ (\ref{WKBbulk}), we recover the
Wannier-Stark ladder \cite{Wannier} for the bulk states energies:
\begin{equation}
E_M=M\Omega.
\label{EMbulk}
\end{equation}
When Eq.\ (\ref{EMbulk}) applies, the function $\varphi_M(k)$ in Eq.
(\ref{phase}) is periodic: $\varphi_M(k)=\varphi_M(k+2\pi)$, thus
integral (\ref{phiMy}) can be taken only from $-\pi$ to $\pi$, because
the integrals along the vertical portions of the integration contour
cancel. In this case, the bulk wave functions are expressed in terms
of the Bessel functions $J$ of an integer order:
$\phi_M(n)=J_{n-M}(2t/\Omega)$.

In all cases, as follows from Eqs.\ (\ref{phiMy}) and (\ref{phase})
with the contours of integration shown in Fig.\ \ref{complex}, the
electron wave functions
\begin{eqnarray}
\phi_M(n)&=&\int\frac{dk}{2\pi}
\exp\left(ink-i\frac{E_Mk+2t\sin k}{\Omega}\right)
\nonumber \\
&=&J_{n-E_M/\Omega}(2t/\Omega)
\label{Bessel}
\end{eqnarray}
are nothing but the Bessel functions of a general order $n-E_M/\Omega$
in the Sommerfeld representation \cite{Nikiforov}. The quantized value
of the energy $E_M$ is determined by the boundary condition
(\ref{n=0}):
\begin{equation}
J_{-E_M/\Omega}(2t/\Omega)=0.
\label{J=0}
\end{equation}
The quantization condition in the form (\ref{J=0}) was found in Ref.\ 
\cite{Fukuyama73}. As shown in Appendix, Ref.\ \cite{Azbel-1990} would
have obtained the same Eq.\ (\ref{J=0}), if mathematical errors were
not made there.

The electron wave functions can be expressed in terms of the Bessel
functions (\ref{Bessel}) only when $\varepsilon_\perp(k)=-2t\cos k$ in
Eqs.\ (\ref{Y}) and (\ref{varphiMy}), which corresponds to the
electron tunneling between the nearest neighboring chains.  Proper
description of the Bechgaard salts requires to take into account
higher harmonics of the transverse dispersion law of electron, such as
$-2t'\cos 2k$, which corresponds to the electron tunneling between the
next-nearest neighboring chains \cite{Zanchi96}.  The WKB method
described in this section is still applicable for an arbitrary
transverse dispersion law $\varepsilon_\perp(k)$, but the wave
functions are not the Bessel functions any more.

\section{Numerical solution and discussion}
\label{sec:numerical}

\begin{figure}
\centerline{\psfig{file=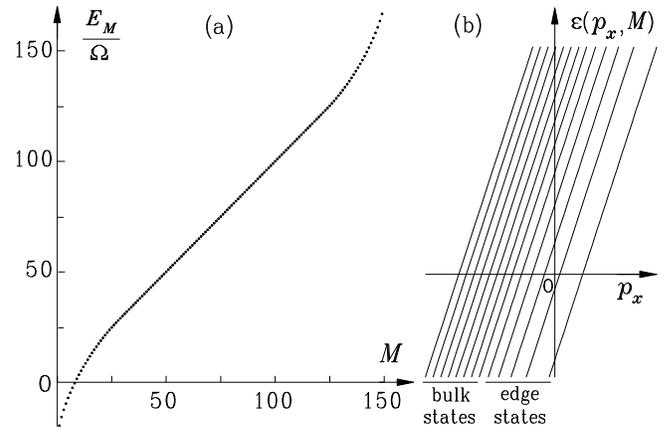,width=\linewidth,angle=-90,clip=}}
\caption{ (a) Eigenenergies $E_M$ of Hamiltonian (\ref{Hmatrix}) found
by numerical diagonalization in the case $n_{\rm max}=150$ and $n_{\rm
edge}=2t/\Omega=25$. (b) Electron dispersion law $\varepsilon(p_x,M)$
(\ref{E}). Only the branches with $M=1,6,11,\ldots,61$ are shown.}
\label{energies}
\end{figure}

To verify the semiclassical results, we solve Eq.\ (\ref{discrete})
for a finite number of chains $n_{\rm max}=150\gg n_{\rm
  edge}=2t/\Omega=25$ by numerical diagonalization of the Hamiltonian
\begin{equation}
\hat{\cal H} =
   \left(
      \begin{array}{cccccc}
        \Omega & -t     & 0      & 0      & \cdots & 0  \\
        -t     &  2\Omega  & -t   & 0     & \cdots & 0  \\
        0      & -t     & 3\Omega & -t    & \cdots & 0  \\
        \vdots & \vdots & \ddots & \ddots & \ddots & \vdots \\
        0      & 0      & \cdots & -t     & (n_{\rm max}-1)\Omega & -t \\
        0      & 0      & \cdots & 0      & -t   & n_{\rm max}\Omega
      \end{array}
   \right).
\label{Hmatrix}
\end{equation}
The quantum probability distributions $|\phi_M(n)|^2$ of two
eigenfunctions of Hamiltonian (\ref{Hmatrix}) (the dots in Fig.\ 
\ref{classical}(b)) agree with the classical probability distributions
(the solid lines in Fig.\ \ref{classical}(b)) of the corresponding
bulk and edge trajectories shown in Fig.\ \ref{classical}(a). The
classical probability distributions are proportional to the square of
Eq.\ (\ref{edge0}) and are equal to $2/Tv_y$, where $v_y=2t\sin
k_*/\hbar\propto 1/\sqrt{(2t)^2-(n'\Omega)^2}$ is the velocity and
$T=\oint dn/v_y$ is the period of classical motion. The numerically
calculated eigenvalues $E_M$ of Hamiltonian (\ref{Hmatrix}), shown in
Fig.\ \ref{energies}(a), agree with the semiclassical energies found
from Eqs.\ (\ref{EMbulk}), (\ref{WKBedge}), and (\ref{Y}) within less
than 1\%. As Fig.\ \ref{energies}(a) demonstrates, the energy levels
are uniformly spaced in the bulk (see Eq.\ (\ref{EMbulk})) with the
energy $\Omega$ (\ref{E_H}) proportional to the magnetic field, but
the spacing is different and not uniform near the edges. In agreement
with Eq.\ (\ref{EMedge}), the spacing of the levels near the edges is
sublinear ($E_M\propto{\rm const}+M^{2/3}$), and the extremal energy
levels with $M=1$ and $M=n_{\rm max}$ are displaced relative to the
linear extrapolation of the bulk law (\ref{EMbulk}) by the amount
$\Delta E\approx\mp2t$.

Transitions between the energy levels $E_M$ in an external ac
electromagnetic field constitute the cyclotron resonance. Because the
penetration depth in metals is short, we expect the energies
(\ref{EMedge}) of the edge states to show up in the surface impedance,
as in conventional metals \cite{Prange68}. The edge states were
neglected in the theory of the cyclotron resonance in Q1D conductors
\cite{Lebed}.

The complete, transverse and longitudinal, dispersion law (\ref{E}) is
shown in Fig.\ \ref{energies}(b). It consists of discrete branches,
each having a continuous linear dispersion in $p_x$. The Fermi momenta
of the branches,
\begin{equation}
  \label{eq:p_F}
  p_F^{(M)}=-E_M/v_F
\end{equation}
are defined as the points where the energy $\varepsilon(p_x,M)$
(\ref{E}) vanishes. The Fermi momenta of the bulk states are spaced
uniformly with the distance $G=\Omega/v_F=ebH/c$, but the spacing is
different and not uniform near the edge. This may have important
consequences for the FISDW state. The FISDW couples the $+P_F$
electrons in the eigenstate $M$ with the $-P_F$ electrons in the
eigenstate $M-N$ \cite{Yakovenko96}. As long as Eq.\ (\ref{EMbulk})
applies, the FISDW wave vector $Q_x=2P_F-NG$ exactly matches the
difference between the Fermi momenta of these states and opens an
energy gap in their spectrum. $N$ branches of the $+P_F$ electrons at
one edge of the crystal and $N$ branches of the $-P_F$ electrons at
the other edge remain gapless, because they have no partners to couple
with \cite{Yakovenko96}. Even though these $2N$ modes are gapless,
electric current is not dissipated, because the modes are chiral, and
the Hall conductivity is quantized: $\sigma_{xy}=2Ne^2/h$, where $h$
is the Planck constant \cite{Yakovenko96}. However, the wave vector
$Q_x=2P_F-NG$ does not match the Fermi momenta (\ref{eq:p_F}) near the
edges, where their spacing is not uniform. Thus, gapless electron
pockets should exist there and cause dissipation in the QHE regime.
The size of the pockets may be reduced if $Q_x$ adjusts to the spacing
of the edge states. The energetics involved in the latter effect
requires a separate study.

\section{Equilibrium currents and magnetization}
\label{sec:currents}

Since the transverse eigenfunctions $\phi_M(n)$ are real, they carry no
electric current across the chains. The current carried by eigenstates
(\ref{psi}) along the chains is
\begin{equation}
j^\pm_{p_x,M}(n)=e\left(\pm v_F-\frac{eA_x(n)}{cm_e}\right)
|\phi^\pm_M(n)|^2,
\label{j}
\end{equation}
where $m_e$ is the electron band mass, and the signs $\pm$ refer to
the $\pm P_F$ electrons. To find the total current $I$ at the chain
$n$, we sum Eq.\ (\ref{j}) over $M$ and integrate over $p_x$ with the
Fermi distribution function (at zero temperature):
\begin{eqnarray}
I(n)&=&2e\sum_{M=1}^{n_{\rm max}}
\int\limits_{-P_F-p_F^{(M)}}^{P_F+p_F^{(M)}}\frac{dP_x}{2\pi\hbar}
\,(\pm v_F)|\phi^\pm_M(n)|^2
\label{In1} \\
&&-\frac{2e^2A_x(n)}{cm_e}\sum_{M=1}^{n_{\rm max}}
\int\limits_{-P_F-p_F^{(M)}}^{P_F+p_F^{(M)}}\frac{dP_x}{2\pi\hbar}
\,|\phi^\pm_M(n)|^2,
\label{In2}
\end{eqnarray}
where the factor 2 comes from the spin of electrons.  It is understood
that the wave functions $\phi^+_M(n)$ and $\phi^-_M(n)$ should be used
when the integration over $P_x$ is close to $+P_F$ and $-P_F$, and
some interpolating functions should be used for the intermediate
values of $P_x$. The result does not depend on the contributions far
from the Fermi surface.

Taking into account that
\begin{equation}
  \label{eq:+=-}
  |\phi^+_M(n)|=|\phi^-_M(n)|, 
\end{equation}
we find that the integral $\int_{-P_F}^{P_F}dP_x$ in Eq.\ (\ref{In1})
vanishes. The only nonzero contribution to this term comes from the
deviations $p_F^{(M)}$ (\ref{eq:p_F}) from the 1D Fermi momenta $\pm
P_F$:
\begin{equation}
  \label{eq:+-EM}
  \frac{e}{\pi\hbar}\sum_{M=1}^{n_{\rm max}}
  -E^+_M|\phi^+_M(n)|^2 + E^-_M|\phi^-_M(n)|^2.
\end{equation}
Taking into account that the eigenfunctions $\phi^+_M(n)$ form a
complete basis of Hamiltonian (\ref{Hmatrix}) and using the relations
$E^+_M=-E^-_M$ and Eq.\ (\ref{eq:+=-}), we find that Eq.\ 
(\ref{eq:+-EM}) can be rewritten as
\begin{equation}
  \label{eq:nHn}
  -\frac{2e}{\pi\hbar}\langle n|\hat{\cal H}|n\rangle
  =-\frac{2e}{\pi\hbar} n\Omega,
\end{equation}
where $\langle n|\hat{\cal H}|n\rangle$ are the diagonal matrix
elements of Hamiltonian (\ref{Hmatrix}).

Using Eq.\ (\ref{eq:+=-}), the term (\ref{In2}) can be written as
\begin{equation}
  \label{eq:Ax}
  \frac{2eA_x(n)}{cm_e}\int_{-P_F}^{P_F}\frac{dP_x}{2\pi\hbar}
  \sum_{M=1}^{n_{\rm max}}|\phi^\pm_M(n)|^2.
\end{equation}
Taking into account the completeness relation \linebreak
$\sum_M|\phi^\pm_M(n)|^2=1$ 
and integrating over $P_x$, we transform Eq.\ (\ref{eq:Ax}) into
\begin{equation}
  \label{eq:nOmega}
  \frac{2eA_x(n)v_F}{c\pi\hbar}=\frac{2en\Omega}{\pi\hbar}.
\end{equation}

The two terms (\ref{eq:nHn}) and (\ref{eq:nOmega}) cancel each other,
so that the total electric current on any chain $n$ is zero:
\begin{equation}
I(n)=0.
\label{zero}
\end{equation}
Because the current vanishes everywhere including the edges, there is
no orbital magnetization (and no de Haas-van Alphen oscillations
proposed in Refs.\ \cite{Azbel87,Azbel-1990}) in a Q1D metal in a
magnetic field. Experimentally, no magnetization was found in the
Bechgaard salts in the metallic state \cite{Chaikin85} (unlike in the
FISDW state, where energy gaps exist in the electron spectrum).

\section{Magnetization in the case of quadratic dispersion law}
\label{sec:magnetization}

In the previous section, we found that orbital magnetization of the
system vanishes identically.  That is a consequence of the linearized
longitudinal energy dispersion law of electrons in our model.
However, for a nonlinear dispersion law, magnetization is not
necessarily zero.  We can crudely estimate the change in the bulk free
energy per one electron at zero temperature generated by an applied
magnetic field, $\Delta F$, in the following way.  $\Delta F$ must
vanish when $H\to0$ and when $t\to 0$.  (When $t=0$, the magnetic
field has no orbital effect on 1D uncoupled chains.)  Because $\Delta
F$ does not depend on the signs of $H$ and $t$, it should be quadratic
in $\Omega=ebHv_F/c$ and $t$ in the lowest order.  To achieve the
dimensionality of energy, we need to divide the expression by a power
of the Fermi energy $\varepsilon_F=P_Fv_F/2$.  In this way, we find
\begin{equation}
\Delta F\sim t^2 \Omega^2/\varepsilon_F^3.
\label{Delta_F}
\end{equation}
Magnetization is obtained by differentiating Eq.\ (\ref{Delta_F}) in
$H$.

It is difficult to calculate $\Delta F$ explicitly in the case of a
weak magnetic field: $\Omega \ll t \ll \varepsilon_F$.  In the
semiclassical (WKB) approximation, the bulk free energy of a Q1D metal
does not depend on the magnetic field, even if the longitudinal
dispersion law is nonlinear, as long as the Fermi surface is open, and
the electron energy spectrum is continuous, not quantized \cite{LAK}.
This result is related to the Bohr-van Leeuwen theorem, which states
that partition function in classical statistical mechanics does not
depend on magnetic field.  Thus, in order to obtain a nonzero $\Delta
F$, it is necessary to go beyond the WKB approximation, which is
difficult.

On the other hand, we can easily calculate $\Delta F$ in case of a
strong magnetic field: $t \ll \Omega \ll \varepsilon_F$, although this
case may not correspond to the Bechgaard salts in realistic magnetic
fields.  In this case, the transverse tunneling amplitude $t$ can be
treated as a small perturbation to the energy spectrum.  The second
order correction to the total energy of the system per one electron at
zero temperature due to a perturbation $V$ is given by the following
expression:
\begin{equation}
  \Delta F^{(2)}=
  \frac{1}{N_e}\sum_{\alpha\:(\varepsilon_\alpha<\varepsilon_F)\atop
    \beta\:(\varepsilon_\beta>\varepsilon_F)} 
  \frac{\langle\alpha|V|\beta\rangle \langle\beta|V|\alpha\rangle}
  {\varepsilon_\alpha-\varepsilon_\beta},
\label{2ndPert}
\end{equation} 
where $N_e$ is the total number of electrons, and the sum is taken
over the energy eigenstates below and above the Fermi energy, labeled
by the indices $\alpha$ and $\beta$, respectively.  Treating the
transverse tunneling amplitude $t$ as the perturbation $V$ and taking
into account that its matrix elements change the longitudinal momentum
$p_x$ by $\pm G$: $\langle\alpha_{k'_x,M'}|V|\beta_{k_x,M}\rangle
=-t\, \delta_{M',M\pm1} \, \delta(k'_x-k_x\mp G)$, we find that the
sum in Eq.\ (\ref{2ndPert}) is restricted to an interval of the width
$G$ in the vicinity of the Fermi momentum:
\begin{equation}
  \delta F^{(2)}=\frac{4 t^2}{\rho_e}\int_{-G}^0\frac{dp_x}{2\pi\hbar}\,
  \frac{1}{\varepsilon_\parallel(p_x)-\varepsilon_\parallel(p_x+G)}.
\label{eq:F2G}
\end{equation}
In Eq.\ (\ref{eq:F2G}), the factor of 4 accounts for the two Fermi
points and two spin orientations, and $\rho_e=4k_F/2\pi$ is the
electron concentration per one chain.  Using the quadratic
longitudinal dispersion law
$\varepsilon_\parallel(p_x)=(P_F+p_x)^2/2m_e$, where $m_e$ is the
effective electron mass, we find:
\begin{eqnarray}
  \Delta F^{(2)}&=&-\frac{4 t^2}{\rho_e}\int_{-G}^0 
  \frac{dp_x}{2\pi\hbar}\,\frac{m_e}{G(P_F+p_x+G/2)}
\nonumber \\
  &=&-\frac{t^2}{\Omega}\ln\left( 
    \frac{P_F+G/2}{P_F-G/2}\right).          
\label{F2}
\end{eqnarray}
Expanding Eq.\ (\ref{F2}) in the small parameter
$G/P_F=\Omega/2\varepsilon_F$ and keeping the first two nonvanishing
terms, we find:
\begin{equation}
  \Delta F^{(2)}=-\frac{t^2}{2\varepsilon_F}
  -\frac{t^2\Omega^2}{96\varepsilon_F^3}+\ldots
\label{eq:F2final}
\end{equation}
The first term in Eq.\ (\ref{eq:F2final}) coincides with the
second-order correction due to the electron tunneling between the
chains in the absence of magnetic field.  Only this,
magnetic-field-independent term is obtained, if the longitudinal
dispersion law is linearized in Eq.\ (\ref{eq:F2G}).  The second term
in Eq.\ (\ref{eq:F2final}) appears due to nonlinearity of the
dispersion law and reproduces the result of dimensional analysis
(\ref{Delta_F}) up to a numerical factor.  Its negative sign indicate
paramagnetism.  However, because the carriers in (TMTSF)$_2$X are
holes with a negative $m_e$, the orbital response would be diamagnetic
in these materials.

\section{Conclusions}
\label{sec:conclusions}

In conclusion, the energy of either a bulk or an edge electron state
in a Q1D metal is the sum of two terms (\ref{E}), one of which has a
continuous spectrum and the other discrete. The discrete part of the
electron energy is determined by the semi-infinite Wannier-Stark
equation (\ref{discrete}).  We have solved the semi-infinite
Wannier-Stark problem semiclassically and numerically.  The WKB
quantization condition (\ref{WKB}) of the electron phase space area
(Fig.\ \ref{classical}(a)) determines the energies of the edge states
with the constant $\gamma=-1/4$ (\ref{WKBedge}) and the bulk states
with $\gamma=0$ (\ref{WKBbulk}).  The energies are spaced uniformly in
the bulk, but not near the edges (see Fig.\ \ref{energies}(a) and
Eqs.\ (\ref{EMbulk}) and (\ref{EMedge})).  These results may be
important for the cyclotron resonance and the QHE in the Bechgaard
salts, as well as the finite-size GaAs-Ga$_{1-x}$Al$_x$As
superlattices in an electric field. We have demonstrated explicitly
that the equilibrium electric currents vanish both at the edges and in
the bulk, so no orbital magnetization is expected in a Q1D metal in a
magnetic field in the approximation of linearized longitudinal energy
dispersion law of electrons.  We have also estimated the magnitude of
orbital magnetization for the quadratic longitudinal dispersion law.

\section*{Acknowledgments}

VMY is grateful to P.~M.~Chaikin, A.~H.~MacDonald, R.~E.~Prange, and
S.~Das Sarma for useful discussions and to also M.~Ya.~Azbel for
drawing attention to Ref.\ \cite{Azbel-1990}. This work was partially
supported by the NSF under Grant DMR--9417451 and by the Packard
Foundation.

\section*{Appendix}

In this Appendix, we use the notation of Ref.\ \cite{Azbel-1990}.
Eq.\ (8) of Ref.\ \cite{Azbel-1990}:
\begin{equation}
\frac12\pi MJ_{\nu}(M)J_{1-\nu}(M)=\sin(\pi\nu),
\label{Azbel}
\end{equation}
can be simplified by using identity (9.1.15) from Ref.\ \cite{Abramowitz}:
\begin{equation}
J_{\nu+1}(M)J_{-\nu}(M)+J_{\nu}(M)J_{-\nu-1}(M)
=-\frac{2\sin(\pi\nu)}{\pi M}.
\label{Wronskian}
\end{equation}
Substituting Eq.\ (\ref{Wronskian}) into Eq.\ (\ref{Azbel}), we find:
\begin{equation}
J_{\nu}(M)[J_{-\nu+1}(M)+J_{-\nu-1}(M)]=-J_{\nu+1}(M)J_{-\nu}(M).
\label{first}
\end{equation}
Using the recurrence relation (9.1.27) of Ref.\ \cite{Abramowitz}:
\begin{equation}
J_{\nu+1}(M)+J_{\nu-1}(M)=\frac{2\nu}{M}J_{\nu}(M),
\label{recurrence}
\end{equation}
in Eq.\ (\ref{first}), we find:
\begin{equation}
J_{-\nu}(M)\left[\frac{2\nu}{M}J_{\nu}(M)-J_{\nu+1}(M)\right]=0.
\label{second}
\end{equation}
Using the recurrence relation (\ref{recurrence}) in Eq.\ 
(\ref{second}) again, we find:
\begin{equation}
J_{-\nu}(M)J_{\nu-1}(M)=0.
\label{product}
\end{equation}
Eq.\ (\ref{product}) is satisfied if either
\begin{equation}
J_{-\nu}(M)=0,
\label{right}
\end{equation}
or
\begin{equation}
J_{\nu-1}(M)=0.
\label{left}
\end{equation}

Eq.\ (\ref{right}) is the same as our energy quantization condition
Eq.\ (\ref{J=0}). Eq.\ (\ref{left}) describes unphysical electron
states located outside of the crystal ($m\leq0$) and should be
discarded.  The two sets of eigenvalues, (\ref{right}) and
(\ref{left}), are completely decoupled and do not repel when cross.
Thus there should be no gaps in Fig.\ 2 of Ref.\ \cite{Azbel-1990}, no
the fractional interference between the two set of the energy levels,
and no diamagnetic oscillations.  Contrary to the explicit
transformation given above, the two sets of eigenvalues come out
coupled via a constant $A$ in Eq.\ (10) of Ref.\ \cite{Azbel-1990}.
We conclude that there must be an error in the ``rather boring
calculations'' mentioned between Eqs.\ (9) and (10) and leading from
Eq.\ (8) to Eq.\ (10) in Ref.\ \cite{Azbel-1990}. The conclusions
following Eq.\ (10) in Ref.\ \cite{Azbel-1990} are invalid because of
the error in this equation.



\begin{references}
\bibitem{TMTSF} TMTSF is tetramethyltetraselenafulvalene and X
represents a monovalent inorganic anion.
\end{references}
\end{document}